\documentclass{article}

\usepackage{microtype}
\usepackage{graphicx}
\usepackage{booktabs} %
\usepackage{multirow}
\usepackage{xspace}
\usepackage{makecell}

\usepackage{caption}
\usepackage{subcaption}

\usepackage{hyperref}

\newcommand{\RLChords}{{\texttt{ReaLchords}}\xspace}

\newcommand{\pastxy}{x_{<t},y_{<t}}
\newcommand{\Pdata}{p_\mathrm{data}}
\newcommand{\Puser}{p_\mathrm{user}}
\newcommand{\Ponline}{\pi}
\newcommand{\Poffline}{\phi}

\usepackage[accepted]{icml2024}

\usepackage{amsmath}
\usepackage{amssymb,amsfonts}
\usepackage{mathtools}
\usepackage{amsthm}

\DeclarePairedDelimiterX{\infdivx}[2]{(}{)}{%
  #1\;\delimsize\|\;#2%
}
\newcommand{\kldiv}{D_{KL}\infdivx}
\DeclareMathOperator*{\E}{\mbox{\Large$\mathbb{E}$}}

\usepackage[capitalize,noabbrev]{cleveref}

\theoremstyle{plain}

\theoremstyle{definition}

\theoremstyle{remark}

\usepackage[textsize=tiny]{todonotes}

\icmltitlerunning{Adaptive Accompaniment with ReaLchords}

\begin{document}

\twocolumn[
\icmltitle{Adaptive Accompaniment with ReaLchords}

\icmlsetsymbol{left}{\#}

\begin{icmlauthorlist}
\icmlauthor{Yusong Wu}{gdm,mila}
\icmlauthor{Tim Cooijmans}{mila}
\icmlauthor{Kyle Kastner}{google}
\icmlauthor{Adam Roberts}{gdm}
\icmlauthor{Ian Simon}{gdm}
\icmlauthor{Alexander Scarlatos}{gdm,umass}
\icmlauthor{Chris Donahue}{gdm,cmu}
\icmlauthor{Cassie Tarakajian}{gdm}
\icmlauthor{Shayegan Omidshafiei}{pgoogle,fieldai}
\icmlauthor{Aaron Courville}{mila,cifar}
\icmlauthor{Pablo Samuel Castro}{gdm,mila}
\icmlauthor{Natasha Jaques}{gdm,uw}
\icmlauthor{Cheng-Zhi Anna Huang}{gdm,mila,cifar}
\end{icmlauthorlist}

\icmlaffiliation{mila}{Mila - Quebec AI Institute, Université de Montréal}
\icmlaffiliation{cifar}{Canada CIFAR AI Chair}
\icmlaffiliation{gdm}{Google DeepMind}
\icmlaffiliation{pgoogle}{Work done while at Google}
\icmlaffiliation{umass}{University of Massachusetts Amherst}
\icmlaffiliation{google}{Google}
\icmlaffiliation{uw}{University of Washington}
\icmlaffiliation{fieldai}{Field AI}
\icmlaffiliation{cmu}{Carnegie Mellon University}

\icmlcorrespondingauthor{Yusong Wu}{wu.yusong@mila.quebec}
\icmlcorrespondingauthor{Cheng-Zhi Anna Huang}{anna.huang@mila.quebec}

\icmlkeywords{Music Generation, Real-time Music Accompaniment, Reinforcement Learning}

\vskip 0.3in
]

\printAffiliationsAndNotice{}  %

\begin{abstract}
Jamming requires coordination, anticipation, and collaborative creativity between musicians.
Current generative models of music produce expressive output but are not able to generate in an \emph{online} manner, meaning simultaneously with other musicians (human or otherwise).
We propose \RLChords, an online generative model for improvising chord accompaniment to user melody.
We start with an online model pretrained by maximum likelihood, and use reinforcement learning to finetune the model for online use.
The finetuning objective leverages both a novel reward model that provides feedback on both harmonic and temporal coherency between melody and chord,
and a divergence term that implements a novel type of distillation from a teacher model that can see the future melody.
Through quantitative experiments and listening tests, we demonstrate that the resulting model adapts well to unfamiliar input and produce fitting accompaniment.
\RLChords opens the door to live jamming,
as well as simultaneous co-creation in other modalities.
\end{abstract}

\begin{figure*}[t]
\centering
\includegraphics[width=\linewidth]{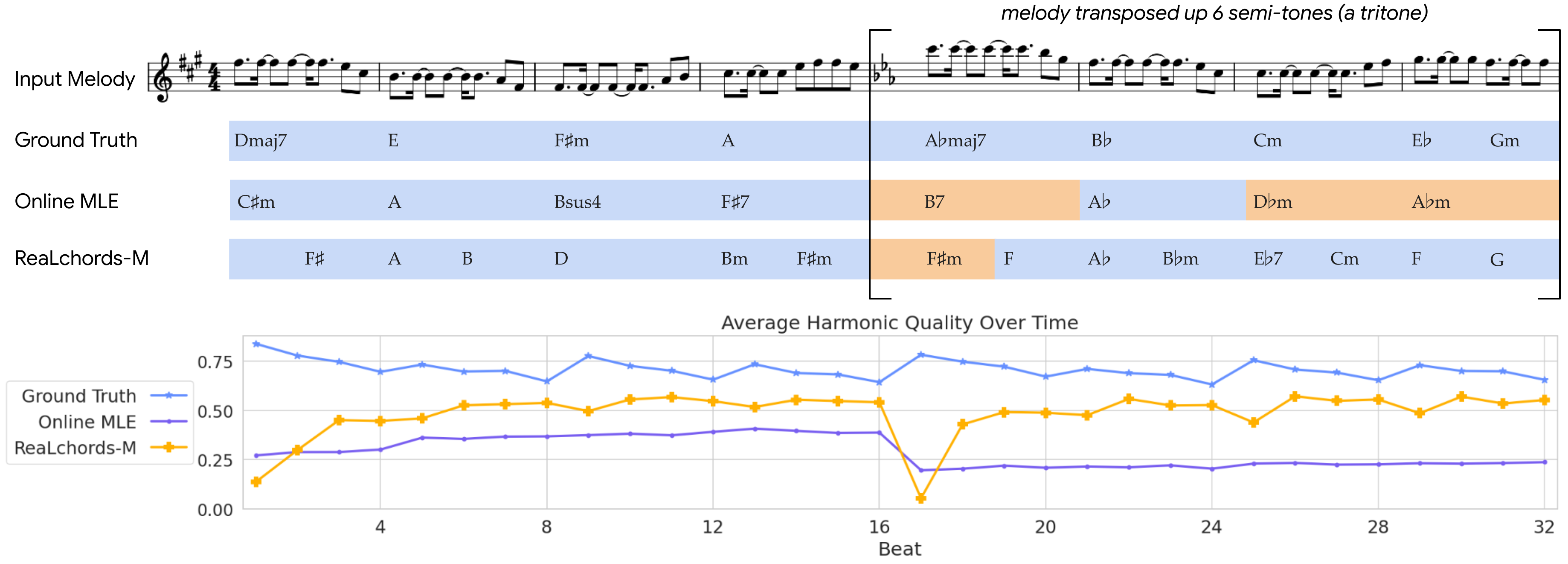}
\caption{Online models finetuned with RL are able to recover from mistakes, while models trained with MLE alone do not.
We take a melody from the test set and midway introduce an abrupt transposition designed to disrupt the accompaniment model (top row).
The Online MLE model predicts a bad chord (B7) and fails to adapt.
\RLChords also predicts a bad chord (F$\sharp$m), but adapts quickly.
Wrong chords highlighted in orange are our own judgment informed by music theory, but the overall pattern is corroborated by an objective measure of harmonic quality, averaged over many trials of this experiment (bottom row).
}
\label{fig:model_generation}
\end{figure*}

\section{Introduction}
\label{submission}
Deep generative models produce realistic, high-quality content, %
and are seeing increasing integration into the creative processes of artists. %
However, such models tend not to be designed for the demands of live scenarios such as interactive improvisation,
which requires anticipation of others' intentions
and adaptation to mistakes, stylistic choices and deliberate exploration.
We focus on music in particular, which is inherently interactive and dynamic and revolves around anticipatory collaboration in ensemble settings.
Most existing models in this space, while capable of creating expressive compositions %
or accompaniments,
are not suited for simultaneous creation,
where adaptation to and alignment with the ongoing musical structure are crucial.

This paper introduces \RLChords, a generative model tailored for online adaptive musical accompaniment. Emulating the spontaneity of live music jamming, \RLChords generates chord accompaniments in response to a stream of monophonic melody notes, adapting on-the-fly to the unfolding musical narrative.
Each chord must be generated without knowing in advance which melody note it will accompany.
This simultaneous interaction imposes a conditional independence assumption on the joint generative process,
that an \emph{online} model must respect.
Moreover, a model must be able to gracefully handle unfamiliar situations and unexpected changes.
Likelihood models, however, suffer from exposure bias due to being trained entirely on ground truth data, and transfer poorly to the online settings where mistakes, imperfections and stylistic differences are common (see Figure~\ref{fig:model_generation} for an example). 

To address this, we use RL finetuning  %
to improve the model with respect to reward models that consider musical coherence (\S\ref{sec:rl_finetuning}).
These reward models see the entire composition and evaluate its musical coherence from various perspectives (\S\ref{sec:reward_model}).
Our setup bears similarities to
RLHF~\citep{ouyang2022training,jaques2020human} and RLAIF~\cite{bai2022constitutional,lee2023rlaif},
however our reward models are trained through self-supervision rather than human labeling.
Finally, we combine RL finetuning with knowledge distillation~\cite{agarwal2023gkd, zhou2023distillspec} in a novel way,
distilling from a teacher that can see the future into a student that cannot, hence forcing anticipation (\S\ref{sec:anchor_model}).

\begin{figure*}[ht]
\centering
\includegraphics[width=\textwidth]{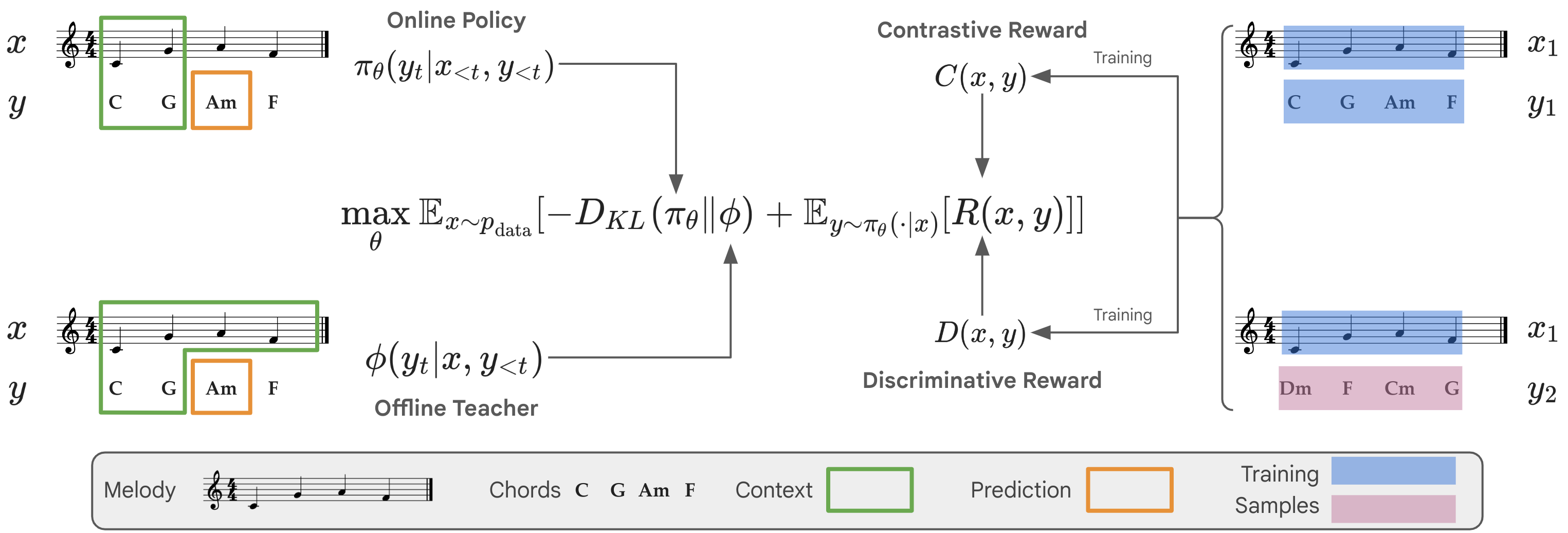}
\caption{\RLChords leverages RL finetuning to learn anticipation and adaptation for online melody-to-chord accompaniment. Initializing from a model $\Ponline_\theta$ pretrained by MLE, the policy generates a complete chord response to a melody from the dataset, each chord being predicted given only previous melody and chords (top left). In contrast, the offline model $\Poffline_\omega$ (also trained by MLE) predicts each chord given the complete melody (bottom left). A KL-divergence penalty distills the predictions of the offline model into the online model, improving its ability to anticipate the future. (Right) The reward stems from an ensemble of multi-scale contrastive and discriminative models that evaluate the musical coherence between melody and chord. The final training objective in \RLChords is a sum of the reward and the distillation loss (center).}
\label{fig:hero_diagram}
\end{figure*}

We develop key algorithmic components (Figure \ref{fig:hero_diagram}) needed to produce an online adaptive accompaniment model that is amenable to interactive use. %
Our contributions and findings are as follows:
\footnote{Listen to audio examples here: \url{https://storage.googleapis.com/realchords/index.html}.}
\begin{itemize}
    \item[$\sharp$] We propose \RLChords, an online accompaniment generation model
    trained by RL finetuning. Figure~\ref{fig:model_generation} shows how \RLChords adapts to out-of-distribution input, a necessary skill for live jamming.
    \item[$\sharp$] We leverage knowledge distillation to learn %
    from a non-causal teacher that can see the future (\S\ref{sec:anchor_model}).
    Distillation greatly improves the quality of the model, as evidenced by the human evaluation shown in Figure~\ref{fig:feedback}.
    \item[$\sharp$] We further employ a novel set of self-supervised reward models to encourage musical coherence and perceptual quality (\S\ref{sec:reward_model}).
    Based on a human listening test, we show that our reward models align closely with human preferences (Figure~\ref{fig:feedback}), despite being trained without human feedback (\S\ref{sec:reward_model}).
    \item[$\sharp$] We demonstrate through a series of controlled experiments that without RL finetuning, models fail to adapt to mistakes and perturbations (Figure~\ref{fig:metrics_per_beat}, \S\ref{sec:adaptation_measure}).
    \item[$\sharp$] Finally, we analyze the behavior of our models in terms of domain-specific metrics (Table~\ref{tab:main_result}, \S\ref{sec:quantitative_results}).
    We find that each component in our RL finetuning methods improves the rhythmic and harmonic quality of generated accompaniments. 
\end{itemize}

\section{Related Work}

\textbf{Adaptive music accompaniment systems} \space\space
In contrast to automatic music generative systems, accompaniment systems often take input (such as melody) from a user, and generate output that is meant to be played in synchrony to complement what the user is playing. Some of these systems are asynchronous, where the user first provides the full melody, and the system generates an accompaniment offline. Examples include MySong~\cite{simon2008mysong}, where a user sings a melody and the system generates chords to accompany them. Most recently, SingSong~\cite{donahue2023singsong} supports a very similar interaction, but generates full-band backing tracks. Both are offline systems.

In contrast, online accompaniment systems need to synchronize with user actions in real-time. Score-following is a special case where the system has the score, the full context of the content of what the musician will play, but still needs to follow along and infer \emph{when} to play their own part. 
Music Plus One~\cite{raphael2010music} adapts its playback speed of an orchestral recording (without the soloist) to a soloist's expressive performance. 
Similarly, Antescofo~\cite{cont2008antescofo} follows where a soloist is in a score and triggers live electronics accordingly.

Generative accompaniment systems or more generally co-creative music systems, not only have to anticipate user actions, they need to learn how to respond. Voyager~\cite{lewis2003too} takes a rule-based approach in how to listen, respond and generate musical material on the fly, while Omax Brothers~\cite{assayag2006omax} recombines what a musician plays on-the-fly as an accompaniment but often requires another computer musician to control when it comes in and what section of material to draw from. ImproteK and later DJazz~\cite{nika2012improtek, nika2017improtek} leverages a shared predefined chord progressions (such as a Jazz Standard) to coordinate the human-machine improvisation. Instead of tight synchronization, Spire Muse~\cite{thelle2021spire} serves as a brainstorming partner which retrieves musical responses that are more or less similar depending on if the user is in a converging or diverging phase of ideation. 

Recent systems based on deep neural networks have emerged. BachDuet~\cite{benetatos2020bachduet} trains an LSTM model using MLE for counterpoint (melody to bassline) accompaniment. SongDriver~\cite{wang2022songdriver} focuses on online melody-to-chord accompaniment, similar to our work. To address exposure bias, SongDriver employs two MLE-trained models: a transformer model that predicts current output based on both current and past outputs, and a conditional random field (CRF) model that predicts current output based on previous context. The CRF model makes online predictions but does not use its own predictions for future contexts; instead, it relies on the transformer model for context.

In contrast, our system \RLChords learns how to respond and in tight synchronization with user melody, by first learning interdependencies between melody and accompaniment from existing songs, and then using RL to tune the models to respond in an adaptive fashion.

\textbf{RL finetuning for generative models} \space\space
Reinforcement learning (RL) finetuning has proven effective in aligning language models with human preferences~\cite{ouyang2022training,jaques2020human} and constraints~\cite{jaques2017sequence}, which are often unaddressed in generative pretraining. In some cases, RL finetuning has been applied to enhance music generation models \cite{jaques2017sequence,jiang2020rl}. Most closely related to our work is RL-Duet~\citep{jiang2020rl},
which considers a similar online generation setting, namely a duet between a user and an agent,
both of them playing each note without knowing what the other will play. Our work provides several contributions over RL-Duet. First, RL-Duet is trained on Bach Chorales, a small dataset of approximately 400 songs following strict rules of counterpoint composition in the style of a particular composer. In contrast, our models are trained on the diverse Hooktheory dataset of 38,000 popular songs from a wide array of artists. To enable effective learning on this scale, we develop novel multiscale contrastive and discriminative reward models, and also propose a new knowledge distillation technique specifically geared toward the online generation setting. Finally, RL-Duet experiments are limited to the setting in which the RL model is primed with the first few ground-truth notes of the accompaniment, an unrealistic assumption for real-time collaborative jamming. As we will show in \S\ref{sec:adaptation_measure}, our methods are able to begin jamming with the user's melody within a few beats, and adapt to sudden perturbations in the key.

Our work is related to the emerging literature on Reinforcement Learning from AI Feedback (RLAIF)~\cite{saleh2020hierarchical, bai2022constitutional, lee2023rlaif}, which mitigates the need for extensive human labeling by utilizing an AI assistant for feedback generation. We use this strategy to finetune online music language models, using an MLE model to obtain a learning signal. Recently, \citet{agarwal2023gkd} have shown that adding a distillation objective between the policy and a larger teacher model during RL finetuning further improves performance. \RLChords employs a novel knowledge distillation objective between the online policy and an offline model which can see future context, %
bridging the gap between online improvisational capabilities and offline musical coherence.

\section{Online Musical Accompaniment}\label{sec:method}

We seek a generative model that can be used for interactive music accompaniment,
where a user plays a melody, and the model simultaneously plays chords to support it.
Accompaniment is a special case of the general setting in which two
agents generate a joint sequence $(x_1,y_1),\ldots,(x_T,y_T)$
in chronological order.
At each step $t$, agents observe the historical material $x_{<t},y_{<t}$,
and simultaneously emit the next pair of tokens $x_t,y_t$.
Simultaneity imposes a conditional independence on the generative process:
\[\Pr(x_t,y_t\mid \pastxy) = \Pr(x_t\mid \pastxy) \Pr(y_t\mid \pastxy).\]
In this general setting, the melody $x$ and chords $y$ interdepend through the conditioning on shared history $x_{<t},y_{<t}$; this corresponds to musicians adapting to each other as they play.
As a first step, we consider the specific setting where the chords do not influence the melody;
now one player leads and the other follows.
We call this accompaniment.

We approach this problem by constructing a model $\pi_\theta$
that generates accompaniment $y$ according to a specific autoregressive process:
\begin{equation}\label{eqn:online}
    \Ponline_\theta(y \mid x) =\prod_t
    \Ponline_\theta(y_t \mid x_{<t}, y_{<t}).
\end{equation}
While our goal at each timestep $t$ is to predict a chord $y_t$ that supports the melody token $x_t$ about to be played, the model's prediction of $y_t$ does not depend on $x_t$.
This is crucial, as it allows the model to be used online as desired.

We train this model in two steps: pretraining on data (\S\ref{sec:mle}),
followed by finetuning using reinforcement learning (\S\ref{sec:rl_finetuning}).
In the rest of this section, we first describe the general approach, and then detail on the components involved (reward models \S\ref{sec:reward_model}, distillation \S\ref{sec:anchor_model}, and regularizations \S\ref{sec:reg_penalties}).

\subsection{Maximum Likelihood Pretraining}\label{sec:mle}

The first step in training $\Ponline_\theta$ is to apply MLE, maximizing the data likelihood
with respect to $\theta$:
\[\max_\theta \E_{x,y\sim \Pdata} \log \Ponline_\theta(y\mid x).\]
The data distribution $\Pdata$ can be interpreted as standing in for $\Puser$: we simulate user play by sampling fixed melodies from the dataset.
This limits our ability to encourage and assess the model's ability to adapt to out-of-distribution melodies. Nevertheless, the model will still encounter out-of-distribution combinations of melodies and chords during inference.

Unfortunately, applying only MLE training to \textbf{online accompaniment model} suffers from exposure bias~\cite{arora2022exposure}: during training, the model is always conditioned on ground-truth context, but this does not occur during inference. Consequently, MLE models struggle to learn two skills required in online accompaniment~\cite{jiang2020rl, jiang2020counterpoint}. First, the model must \emph{anticipate} what the user is going to play, in order to ensure that its own output agrees with that of the user. Second, the model must be able to \emph{adapt} to and recover from unknown input, due to its own mistakes or those of the user, due to misanticipation, or due to user idiosyncrasies.

As a concrete example, Figure~\ref{fig:model_generation} shows a failure mode of the online MLE model. The model fails to adequately anticipate future inputs, leading to exposure bias and error accumulation due to a distribution mismatch between training and inference. Whenever the first few time-steps of output do not fit with the melody input stream the model will continue its chord progression, \textit{ignoring the input}.

\subsection{Finetuning using Reinforcement Learning}\label{sec:rl_finetuning}

Similar challenges are encountered in imitation learning~\cite{ross2010efficient},
where policies trained by MLE to reproduce expert demonstrations are brittle, and fail to transfer to the real environment (see e.g. \citet{reddy2019sqil}). 
A rich history of work has demonstrated Reinforcement Learning (RL) finetuning to be an effective remedy.

We begin by initializing the weights of our RL policy $\Ponline_\theta$ with those of the pretrained online MLE model. As in eq.~\ref{eqn:online}, at timestep $t$, the policy predicts action probability distribution $a_t=y_t$ given state $s_t=(x_{<t}, y_{<t})$.
Then, we adopt an RL finetuning methodology similar to the popular RLHF (RL from Human Feedback) framework used for language models \cite{ouyang2022training,jaques2020human}.
Namely, in addition to maximizing RL rewards $R(x,y)$, we minimize KL-divergence from a pretrained MLE anchor model $\Poffline_\omega(y|x)$ parameterized by $\omega$, as proposed in \citet{jaques2017sequence}. Let $x$ and $y$ represent the full melody and chord sequence, each consisting of several tokens (i.e. the full \textit{trajectory}). This gives us the KL-regularized RL objective:

\begin{equation}\label{eqn:objective}
\max_{\theta} \E_{\substack{~\\x\sim\Pdata\\y \sim \Ponline_\theta(\cdot \mid x )}}
R(x,y) - \beta \kldiv{\Ponline_\theta(\;\cdot \mid x)}{\Poffline_\omega(\;\cdot \mid x)}.
\end{equation}
To evaluate \eqref{eqn:objective},
we sample a batch of melodies $x$ from the training set, then use the current policy $\Ponline_\theta$ according to \eqref{eqn:online} to generate a batch of corresponding harmonies $y$ (Figure~\ref{fig:hero_diagram}, top left).
We then evaluate the resulting batch of compositions $(x,y)$
according to reward models (\S\ref{sec:reward_model})
and regularizers (\S\ref{sec:reg_penalties})
to obtain the reward $R(x,y)$ (Figure~\ref{fig:hero_diagram}, top and bottom right).
Additionally, we measure $\Poffline_\omega(y\mid x)$ under the offline model $\Poffline_\omega$ (\S\ref{sec:anchor_model}) in order to compute the KL term (Figure~\ref{fig:hero_diagram}, bottom left).
Finally, we update the model according to \eqref{eqn:objective},
using REINFORCE with a separate value model serves as baseline estimation for improved stability~\citep{lee2023rlaif, agarwal2023gkd}. The separate value model is also initialized from pretrained online MLE model, and is trained to estimate the total return. We use mean square error between the estimated return and total return as objective to train the value model.

Unlike in RLHF~\cite{ouyang2022training} and RLAIF~\cite{bai2022constitutional}, our reward models are not trained on preference labels from either human or machine labelers.
Instead, they are trained using positive and negative melody-chord pairs constructed from a dataset (see Figure~\ref{fig:hero_diagram}, \S\ref{sec:reward_model}).
Nevertheless, a listening test (\S\ref{sec:listening_test}) shows that our reward models align well with human preferences, as shown in Figure~\ref{fig:feedback}.

\subsection{Reward Models}\label{sec:reward_model}

We develop a novel ensemble of reward models that evaluates the coherence between input (melody) and output (chord) tracks.
We implement two types of coherence evaluation reward models, contrastive and discriminative, each with different inductive biases.
Reward model training and architectural details can be found in Appendix \S\ref{sec:appendix_reward_models} and \S\ref{sec:appendix_reward_model_performance}.

The \textbf{contrastive reward model} consists of a melody encoder and a chord encoder, which respectively map the melody $x$ and chord $y$ to embedding vectors $E_x,E_y$. The encoders are trained in an unsupervised manner using InfoNCE loss~\cite{oord2018representation, radford2021learning} applied to positive and negative samples created from the dataset. As shown in Figure \ref{fig:hero_diagram}, the positive pairs are defined as the melody-chord pairs from the same song, and the negative pairs are created by randomly pairing melody and chord from different songs.
The InfoNCE loss essentially maximizes the cosine similarity for positive pairs, and minimizes the cosine similarity for negative pairs. 
The reward for a given pair $x,y$ is the cosine similarity of $E_x$ and $E_y$.

The \textbf{discriminative reward model} looks at the entire generated pair $(x,y)$ as a whole.
This model is trained in an unsupervised manner to discriminate between ``real'' melody-chord pairs and randomly paired melodies and chords. Each training batch case provides a set of positive pair and, by combining its melody with the chords from another randomly chosen batch case, a negative pair.
Once trained, the model produces a probability of $(x,y)$ being ``real`` that directly serves as the reward.

Due to the bottleneck on the embedding vectors $E_x,E_y$, the contrastive models focus on global coherence. The discriminative models on the other hand are able to evaluate temporally local coherence. Indeed, our experiments in \S\ref{sec:quantitative_results} show that contrastive reward models promote mainly harmonic quality whereas discriminative reward models encourage mainly synchronization. 

While these reward models are effective, we find that they can be overly harsh on temporally localized incompatibilities, such as short-lived mistakes that are quickly resolved.
To mitigate this and improve temporal credit assignment, we further propose to use an ensemble of \emph{multi-scale} variants that evaluate isolated fragments without being influenced by distant mistakes.
We train multiple contrastive and discriminative reward models to judge fragments of reduced lengths ($\{\frac{1}{2}, \frac{1}{4}, \frac{1}{8}, \frac{1}{16} \}$ of the maximum sequence length 256). During finetuning, we apply these models to sliding windows (50\% overlap) of the example.

\subsection{Distillation from Offline Teacher Model}
\label{sec:anchor_model}

As stated in \S\ref{sec:rl_finetuning}, during RL finetuning we penalize KL-divergence from a model pretrained on the data distribution to ensure the model maintains realistic outputs while maximizing rewards \cite{jaques2017sequence}.
However, unlike in typical RL finetuning, the online MLE model with which our policy is initialized suffers from a lack of robustness to out-of-distribution data, and as such is not an ideal anchor for use with the KL-regularization term.
\citet{agarwal2023gkd} demonstrated how the KL penalty can be used not just to avoid diverging from a checkpoint, but also to distill knowledge from a larger teacher model.
We take this idea one step further and distill knowledge from an \emph{offline} model that can see the future of the melody.

The offline model $\Poffline$ is trained with MLE to autoregressively predict chords given the full melody $x$:
\begin{equation}
 \Poffline_\omega(y\mid x) = \prod_t \Poffline_\omega(y_{t} \mid x, y_{<t}).
\end{equation}
In traditional knowledge distillation, ground truth data is used to obtain the predictions of both the teacher and student models, and a KL loss is then applied to bring the student's predictions closer to the teacher's. Here, it is instead evaluated on samples generated by the current policy. This is a special case of \emph{on-policy knowledge distillation}~\cite{agarwal2023gkd,zhou2023distillspec}, which in general allows any mixture of ground truth data, student samples and teacher samples.
We tested various on-policy knowledge distillation schedules and found it works best when driven by the student (\S\ref{sec:quantitative_results}). Thus, during RL finetuning we only train on outputs from the student. %

\subsection{Regularization Penalties}\label{sec:reg_penalties}

RL finetuning can lead to pathological behavior, such as repetitions and mode collapse~\cite{jaques2017sequence,jiang2020rl}.
We introduce three regularization penalties to discourage specific failure modes: \textbf{Repetition}: Inspired by repetition penalties used for training language models (as in \citet{saleh2020hierarchical,jaques2020human}), we impose a penalty for chords that are held for too long. %
\textbf{Silences}: We impose a penalty for silences beyond the beginning of a phrase.  
\textbf{Ending early}: A penalty is imposed for early end-of-sequence (EOS) tokens.
See \S\ref{sec:appendix_penalties} for an ablation that shows the need of these penalties.

\begin{figure}[t!]%
\captionsetup{justification=justified,singlelinecheck=false}
\centering
\begin{subfigure}[t]{0.9\linewidth}
    \includegraphics[width=\linewidth]{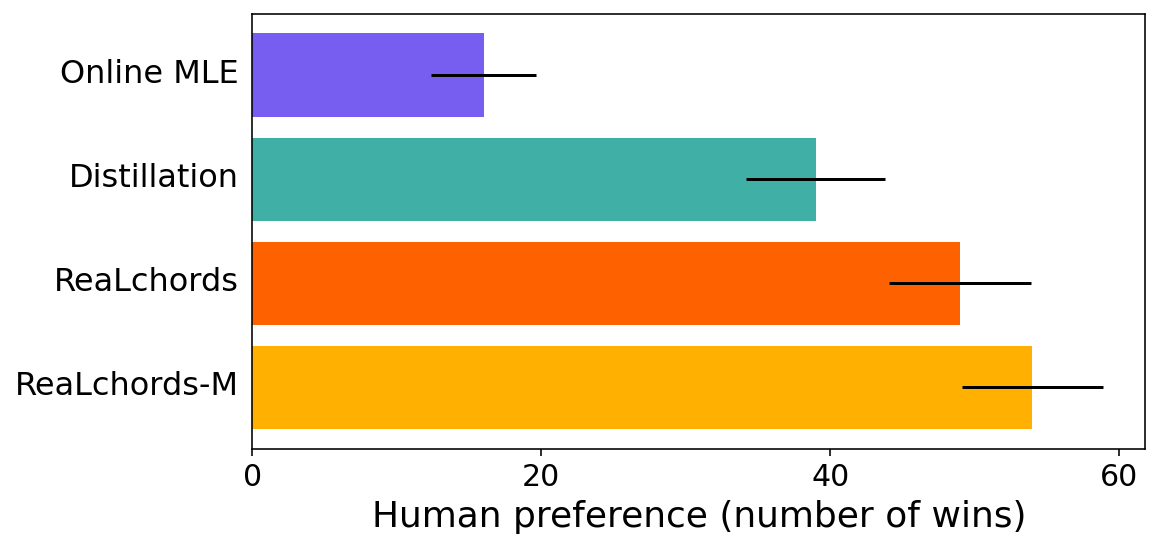}
\end{subfigure}
\vspace{0.1in}
\begin{subfigure}[t]{0.9\linewidth}
    \includegraphics[width=\linewidth]{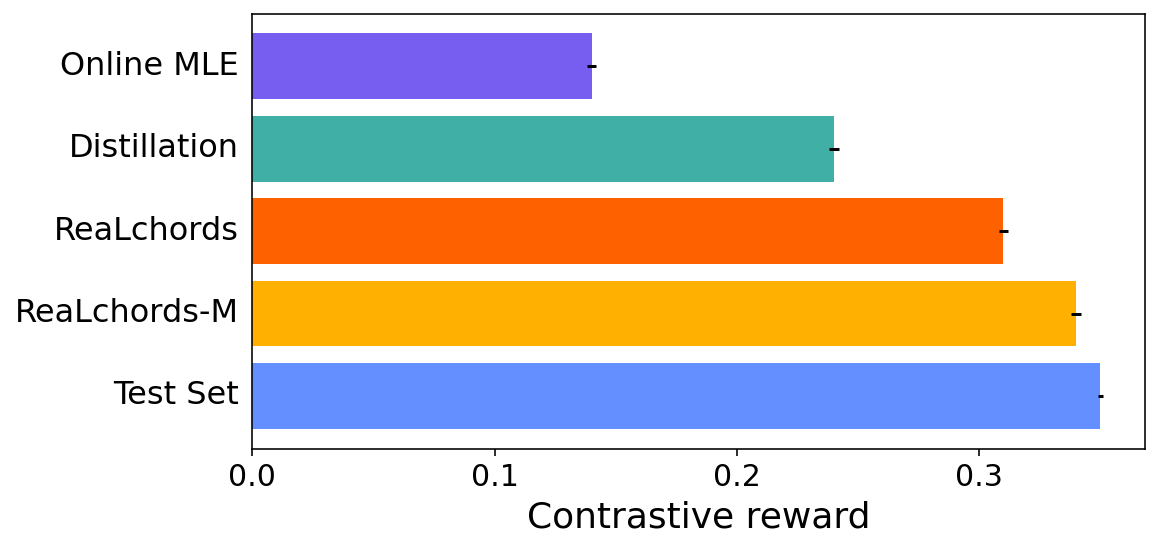}
\end{subfigure}
\vspace{0.1in}
\begin{subfigure}[t]{0.9\linewidth}
    \includegraphics[width=\linewidth]{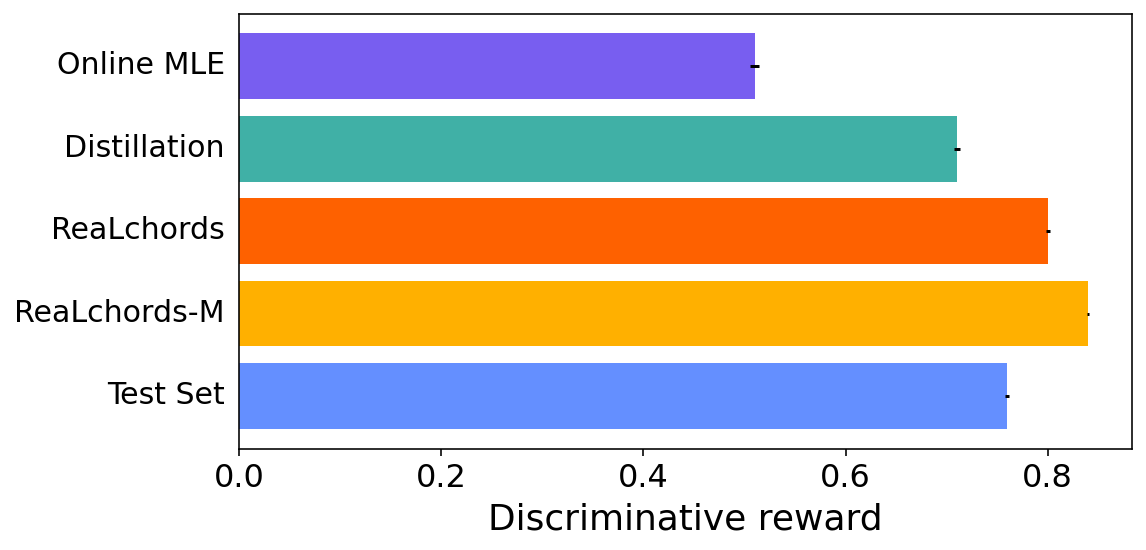}
\end{subfigure}%
\vspace{-0.1in}
\caption{
Our reward models are aligned with human preferences.
We carried out a listening test (\S\ref{sec:listening_test}) to evaluate the quality of our models.
The online MLE model performs poorly, but is greatly improved by distillation from the offline MLE model.
Our proposed systems \RLChords and \RLChords-M improve further thanks to RL finetuning.
The rewards given by both the contrastive and discriminative reward models are strongly correlated with human evaluations.
}
\label{fig:feedback}
\end{figure}

\section{Dataset}

We train our models on an updated version of the Hooktheory dataset~\cite{donahue2022melody}, which comprises crowd-sourced analyses of monophonic melodies and chords from recordings and now contains 38K melody-chord pairs. We adopt a frame-based representation where time is quantized to sixteenth notes, where each frame is a discrete index. We set a maximum sequence length of 256 for $x$ and $y$. We augment the data by randomly transposing up or down by up to 6 semitones. $20\%$ of the data is held out and divided equally into validation and test sets. We develop on the validation set and report the test set results in the paper. Please refer to \S\ref{sec:appendix_dataset} for details on the dataset and data representation.

\begin{table*}[]
\caption{Effect of RL finetuning with our reward models and knowledge distillation on harmonic, synchronization and rhythmic diversity metrics.
Each number is an average over a large number of accompaniments to test set melodies. In row 2-7, we report confidence interval (95\%) of metric values over 3 RL finetunings, each with different random seeds.}
\label{tab:main_result}
\vskip 0.15in
\begin{center}
\begin{small}
\begin{tabular}{@{}lccc@{}}
\toprule
Models                      & \multicolumn{1}{c}{\begin{tabular}[c]{@{}c@{}}Harmony \\ Note in Chord $\uparrow$, \%\end{tabular}} & \multicolumn{1}{c}{\begin{tabular}[c]{@{}c@{}}Synchronization \\ $\Delta$ Chord-Note Onset Interval $\downarrow$, $\times 10^{-3}$ \end{tabular}} & \multicolumn{1}{c}{\begin{tabular}[c]{@{}c@{}}Rhythm Diversity \\ Chord Length Entropy $\uparrow$\end{tabular}} \\ \midrule
Online MLE                  & 36.99                                                                                              & 14.23                                                                                                            & 2.21                                                                                                  \\ \midrule
Knowledge Distillation (KD) & 46.38 $\pm$ 0.45                                                                                               & 14.39 $\pm$ 1.25                                                                                                            & \textbf{1.80} $\pm$ 0.04                                                                                     \\
Contrastive (C)             & 47.22 $\pm$  1.22                                                                                           & 16.61 $\pm$ 4.78                                                                                                           & 1.80  $\pm$  0.29                                                                                              \\
Discriminative (D)          & 44.29 $\pm$ 0.99                                                                                              & \textbf{12.34} $\pm$ 7.95                                                                                                            & 1.70  $\pm$ 0.06                                                                                               \\
C + D                       & 46.12 $\pm$ 0.95                                                                                            & 12.86 $\pm$ 6.17                                                                                                & 1.56 $\pm$ 0.05                                                                                                \\
\RLChords                 & 48.17 $\pm$ 0.27                                                                                              & 16.09 $\pm$ 3.63                                                                                                          & 1.35 $\pm$ 0.30                                                                                                \\
\RLChords-M                  & \textbf{54.29} $\pm$ 1.55                                                                                   & 17.17 $\pm$ 4.86                                                                                                          & 1.66 $\pm$ 0.20                                                                                                \\ \midrule
Offline MLE                 & 63.73                                                                                              & \phantom{0}9.85                                                                                                             & 1.90                                                                                                  \\
Test set                    & 70.94                                                                                              & \phantom{0}0.00                                                                                                                & 2.19                                                                                                  \\ \bottomrule
\end{tabular}
\end{small}
\end{center}
\end{table*}

\section{Experiments}\label{sec:experiments}

Is the system capable of producing accompaniments of high musical quality?
How swiftly can the system adjust to unfamiliar situations?
We address these questions from three directions.

To directly assess musical quality, we conduct a human listening test using samples generated from the models (\S\ref{sec:listening_test}).
We demonstrate adaptation through several controlled generation experiments, tracking the quality of the accompaniment over time (explained in \S\ref{sec:adaptation_measure}).
Finally, we evaluate the system using heuristic metrics to assess the quality of compositions generated in response to melodies in the test set (detailed in \S\ref{sec:quantitative_results}).

The following systems are compared in our experiments:

\textbf{MLE baselines} \space\space
The \textbf{Online MLE} model trained to predict $y_t\mid \pastxy$ without seeing $x_t$ (\S\ref{sec:mle}).
The \textbf{Offline MLE} model that sees the full input $x$ and is used as a teacher for knowledge distillation (\S\ref{sec:anchor_model}).

\textbf{Our proposals} \space\space These models are trained with both contrastive and discriminative rewards, as well as regularization and knowledge distillation.
\RLChords incorporates the global reward models,
whereas \RLChords-M incorporates the multi-scale variants of both reward models.

\textbf{Ablations} \space\space
The model \textbf{KD}, trained with only knowledge distillation and regularization.
Two models trained by MLE and then finetuned using either only \textbf{Contrastive~(C)} reward or only \textbf{Discriminative~(D)} reward, with regularization and KL divergence to the MLE checkpoint.
A model \textbf{C+D} using both contrastive and discriminative reward, with regularization and KL divergence to the online MLE checkpoint.

\subsection{
Human and Machine Evaluation on Musicality
}\label{sec:listening_test}

Any measure of the quality of a musical piece must ultimately be grounded in human preferences.
We carry out a listening test to evaluate four systems:
the Online MLE baseline, KD, \RLChords and \RLChords-M.
In the listening test, participants are presented with 8-second audio clips from two different systems, and asked to rate which one sounded more musical, on a 5-point Likert scale. We recruited ten musicians, and collected 192 ratings with each system involved in 96 pairwise comparisons (see \S\ref{sec:appendix_human} for more details). 

Figure~\ref{fig:feedback} (top) shows the number of wins for each system. We ran a Kruskal-Wallis H test and confirmed that there are statistically significant pairs among the permutations. According to a post-hoc analysis using the Wilcoxon signed-rank test with Bonferroni correction (with p$<$0.05/6 as there are 6 pairs of systems), we found the following statistically significant results: All systems outperformed the Online MLE baseline. Also, the fully-fledged systems \RLChords and \RLChords-M outperformed distillation alone (KD). While \RLChords-M appears to outperform \RLChords, this comparison is not significant.

Overall, the results from the listening test show that distillation alone (KD) accounts for a large improvement in perceptual quality.
The reward models agree with this assessment, even though KD does not directly optimize for these rewards.
In general, we find that the rewards given by our self-supervised reward models (Figure~\ref{fig:feedback}, middle and bottom) correlate strongly with human preferences, which justifies their use in lieu of human feedback.

\subsection{Quantitative Metrics}

In line with prior research~\cite{jiang2020rl, yang2020evaluation, fang2020bach}, we introduce quantitative metrics to evaluate the quality of accompaniments:

\textbf{Harmonic quality} \space\space We measure harmonic quality by the note-in-chord ratio, which is the amount of time that the melody's pitch class occurs in the chord. For example, if the melody token $x_t$ is a C, and the chord $y_t$ is F minor, then the note-in-chord ratio as time $t$ equals 1. We average this metric across time $t$ and across all compositions $x,y$ generated, to obtain the overall note-in-chord ratio for the model in question.

\textbf{Synchronization} \space\space To gauge temporal synchronization between melody and chord progression, we look at \emph{chord-to-note onset interval}, which is the length of time between the onset of a chord and the onset of the nearest preceding melody note.
The synchronization of a model can be judged by comparing this quantity's distribution on the test set versus on the output of the model.
Whereas \citet{jiang2020rl} compare averages of this quantity, we propose to compare the full distributions using Earth Mover's Distance (EMD) on histograms of chord-to-note onset intervals.

\textbf{Rhythmic diversity} \space\space
We examine the distribution of durations of generated chords
to assess overall rhythmic behavior. The entropy of this distribution measures rhythmic diversity.

\subsection{Quantitative Evaluation Results}\label{sec:quantitative_results}

We evaluate each model based on a large number of accompaniments to test set melodies.
The average metrics are reported in Table~\ref{tab:main_result}.

\textbf{MLE baselines} \space\space
The behavior of Offline MLE is closest to that of the test set, as is expected due to its ability to see the future input.
Online MLE exhibits poor harmonic and temporal coordination with the melody,
which suggests that it produces chords without paying attention to the melody.

\textbf{Distillation} \space\space On-policy knowledge distillation (KD in Table~\ref{tab:main_result}) significantly enhances online generation, particularly with regard to harmony and synchronization.
Distillation from the offline teacher will suppress the probability of chords that match poorly with the future melody,
forcing the student to anticipate the immediate future.
The student (KD) learns to produce outputs that align better with the input context while retaining a causal conditioning structure.

\begin{figure}[t!]
\captionsetup{justification=justified,singlelinecheck=false}%
    \begin{center}%
    \begin{subfigure}{\linewidth}%
    \caption{Accompaniment quality when primed with ground truth.}
    \includegraphics[width=\linewidth]{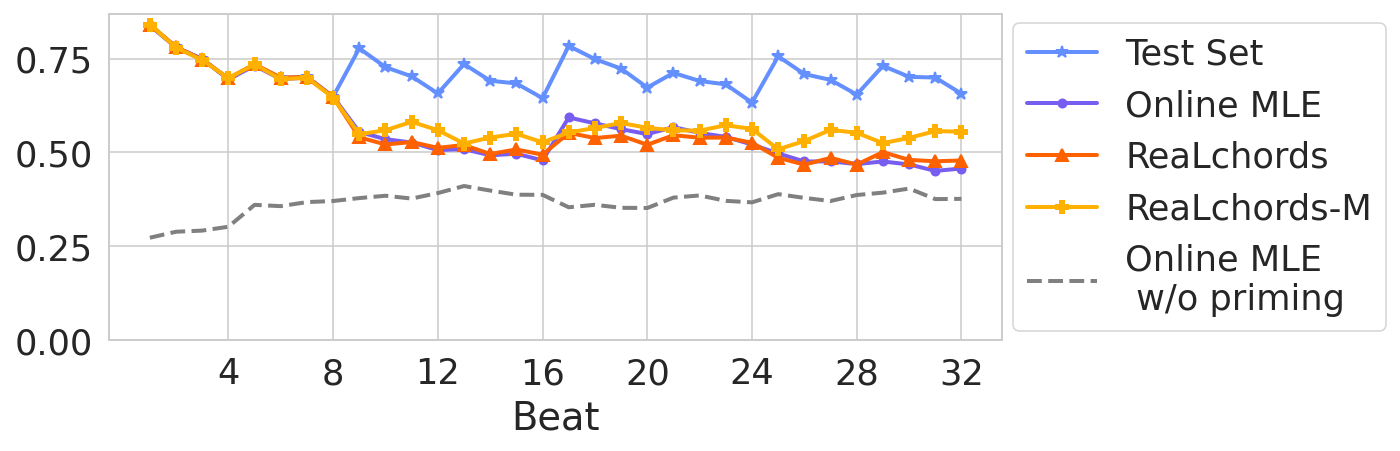}
    \label{fig:priming}
    \end{subfigure}%
    \vspace{-1ex}
    \\
    \begin{subfigure}{\linewidth}%
    \caption{Accompaniment quality after a cold start.}
    \includegraphics[width=\linewidth]{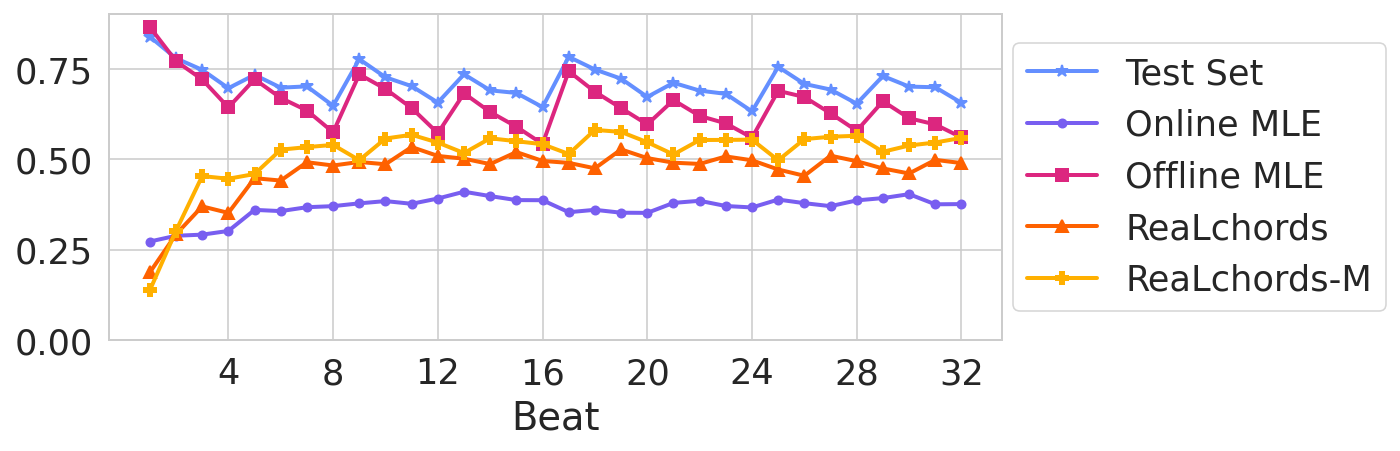}
    \label{fig:original}
    \end{subfigure}%
    \vspace{-1ex}
    \\
    \begin{subfigure}{\linewidth}%
    \caption{Accompaniment quality when perturbed midway.}
    \includegraphics[width=\linewidth]{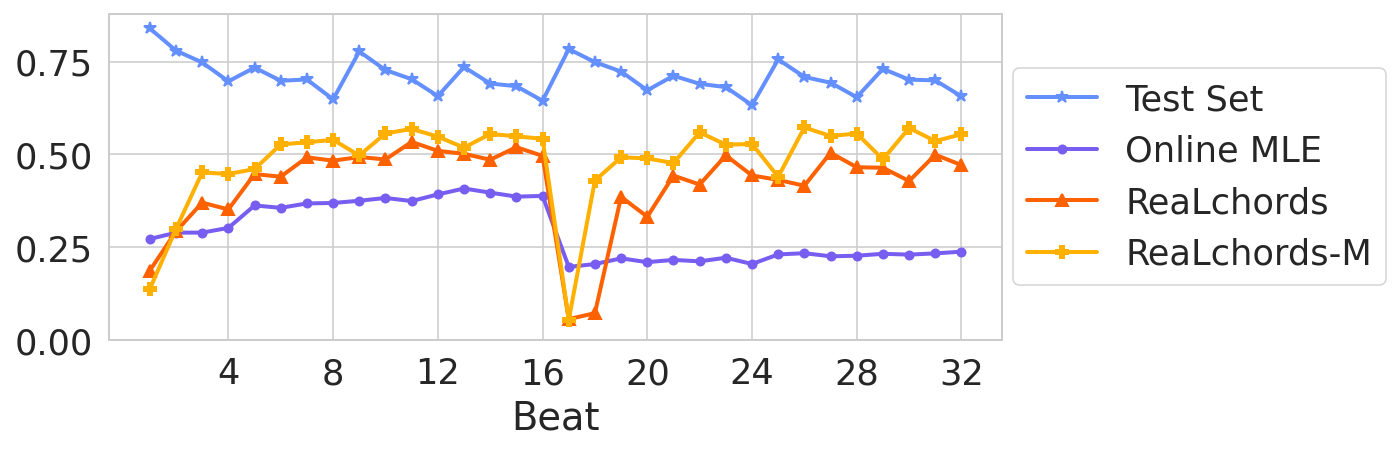}
    \label{fig:transposition}
    \end{subfigure}%
    \vskip -0.1in%
    \caption{Comparing the quality of overall accompaniment as a function of the number of beats generated, in three scenarios of increasing difficulty (\S\ref{sec:adaptation_measure}). Quality is measured by note-in-chord ratio. (a) Priming the online model with ground-truth context (8 beats in this case) results in comparable performance between models. (b-c) \RLChords and \RLChords-M recover from cold starts and perturbation, while the online model does not. }
    \label{fig:metrics_per_beat}
        \end{center}
        \vskip -0.2in
\end{figure}

\textbf{Reward models} \space\space Training with contrastive and discriminative reward models, individually (C, D) and combined (C + D), shows distinct improvements in harmony and synchronization.
The use of the contrastive reward model (C) improves more on harmony, presumably because it compresses the entire melody and the entire accompaniment separately and merges them only in the final cosine similarity. The use of the discriminative reward model (D) improves more on synchronization, while having worse harmony. This is expected, as the classification is biased towards direct comparison. 
We further examine the bias of different reward models by plotting reward values against harmonic perturbations, where varying portions of chords in the test set are replaced with random alternatives. As shown in Figure~\ref{fig:reward_perturb} in \S\ref{sec:appendix_reward_model_performance}, the contrastive model is more sensitive to harmonic perturbations.

The blend of both rewards in \RLChords offers enhancement in each metrics. Additional experiments applying RL fine-tuning with ensembles of the same type of reward models show similar metric improvements, as presented in Table~\ref{tab:ensemble_reward} in \S\ref{sec:appendix_ensemble_reward}. This suggests that the observed metric enhancements may result from both the combined biases and the ensemble of reward models.

\textbf{Combining rewards with distillation} \space\space
Integrating both reward models with knowledge distillation yields better harmony but less rhythmic diversity (\RLChords in Table~\ref{tab:main_result}). This indicates a tendency of the model to opt for `safer' chord progressions, presumably resulting from satisfying both reward maximization and knowledge distillation. This is further validated in Figure~\ref{fig:histograms} in Appendix \S\ref{sec:appendix_histograms} where we visualize the chord length histograms and find that this model tends to hold chords for 2 or 4 beats.

\textbf{Multi-scale reward models} \space\space
\RLChords-M further improves harmonic quality
thanks to the locally isolated rewards from the multi-scale variants of our reward models.
This aligns well with the findings in Figure~\ref{fig:feedback}.

\subsection{Adaptation Dynamics}\label{sec:adaptation_measure}

To study the temporal dynamics of model adaptation to unknown input, we measure accompaniment quality as a function of number of beats generated. A beat is 4 frames with a total of a quarter note length.
We compare among the dataset and four models: Online MLE, Offline MLE, \RLChords and \RLChords-M.
We report harmonic quality in terms of note-in-chord ratio.
In all experiments, we draw melodies from the test set, and let the models generate accompaniment.
However, we consider three different scenarios with different interventions: priming, cold-start, and perturbation.

\textbf{Priming} (Figure~\ref{fig:priming}):
We start with the setting of RL-Duet~\cite{jiang2020rl},
where the models are primed with several beats of ground truth chords before generating their own chords.
This avoids the cold-start problem of predicting a chord without knowing anything about the melody,
and gives an indication of the model's ability to anticipate what happens next without having to first adapt to what happened previously.
Behavior is similar across the models.
For reference, we also plot the cold-start behavior of Online MLE (without priming), which is significantly worse.
We argue that, as a benchmark for online accompaniment, primed generation is unnatural and too facile.

\textbf{Cold start} (Figure~\ref{fig:original}):
We now proceed to the cold-start setting, which is more natural and more difficult.
Here, models predict chords immediately and have to adapt to the resulting melody-chord combinations, which are usually wrong and outside of the data distribution.
The Online MLE%
struggles to adapt to its own past chords, and never gets close to the primed behavior.
\RLChords and \RLChords-M quickly overcome their own mistakes and play as well as if they were primed.

\textbf{Perturbation} (Figure~\ref{fig:transposition}):
Finally, we introduce a deliberate perturbation in the middle of the generation process, to demonstrate the ability of our systems to recover from serious errors.
We transpose the melody up by a tritone (6 semitones) at beat 17, resulting in both an out-of-distribution melody and almost guaranteeing that the next chord is a poor fit.
This is similar to the push test in legged robotics.
The Online MLE fails the test: it exhibits a drop in harmonic quality and never recovers.
\RLChords and \RLChords-M quickly adapt to the new key and recover their previous performance.

Overall, these results confirm that Online MLE suffers from exposure bias due to only being trained on ground-truth data.
This brittleness, or inability to produce reasonable output given out-of-distribution (OOD) input not covered by the training data, is similar to the failures exhibited by imitation learning or behavior cloning methods in traditional RL contexts \cite{reddy2019sqil}, which also rely purely on supervised learning.
Our systems \RLChords and \RLChords-M quickly recover from both cold-start situations and mid-song disturbances,
which highlights their ability to follow along with a user as they explore ideas. Thus, \RLChords solves a critical requirement for an online accompaniment system, which will have to accompany a diverse range of human users who are likely to play novel melodies not covered by the training data and change what they play midway.

Finally, we find an interesting emergent behavior due to RL finetuning,
where models hold off on playing chords initially, preferring instead to wait for more information about the melody.
This \emph{wait and see} behavior is also visible in Figure~\ref{fig:model_generation}, and examined further in Appendix \S\ref{sec:appendix_silence}.
Similar behavior occurs in human performers, who often wait for several bars when improvising with an unfamiliar player.

\section{Conclusion}

We proposed \RLChords, an online generative model that improvise simultaneous chord accompaniment in response to melody input. \RLChords leverages RL finetuning with multi-scale contrastive and discriminative reward models and employs a novel offline-to-online knowledge distillation technique. Throughout our experiments, we show that \RLChords accompanies with good harmony and synchronization, while effectively adapting to mistakes and perturbations. We also show that listeners preferred \RLChords over the online MLE baseline and distillation-only models, while validating the proposed reward models align with human judgement. \RLChords enables an exciting path towards an effective and engaging real-time, interactive music accompaniment system.

\section*{Acknowledgement}
We would like to express our gratitude to the following individuals for their insightful discussions and valuable advice on the project: Jesse Engel, Ethan Manilow, Antoine Caillon, Laura Graesser, Athul Jacob, Kory Mathewson, Max Schwarzer, Evgenii Nikishin, Zhixuan Lin, Dinghuai Zhang, Michael Noukhovitch, Ke Chen, Yuxuan Wu, Yi Deng. We also like to thank the individuals who designed and built the RL training infrastructure used in this paper: Léonard Hussenot, Johan Ferret, Robert Dadashi, Geoffrey Cideron, Alexis Jacq, Sabela Ramos, Piotr Stanczyk, Sertan Girgin, Danila Sinopalnikov, Amélie Héliou, Bobak Shahriari, Bilal Piot, Matt Hoffmann, Nikola Momchev, and Olivier Bachem.

\section*{Impact Statement}

While this work is developed in the domain of music, the interactive dynamic of simultaneous musical accompaniment has broader societal implications beyond the realm of music generation. 
This research could also influence the way AI is perceived in creative fields, reinforcing the potential of AI as a collaborative tool rather than a replacement for human creativity. %

\bibliography{ref}
\bibliographystyle{icml2024}

\newpage
\appendix
\onecolumn

\section{Wait and See Behavior}\label{sec:appendix_silence}

\begin{figure}[t]
    \begin{center}
    \includegraphics[width=\linewidth]{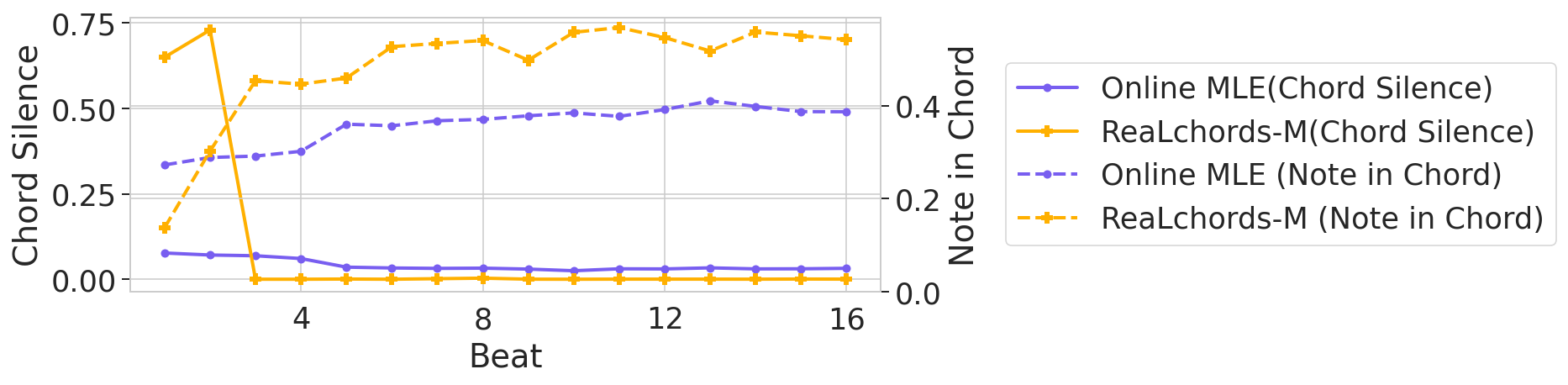}
    \caption{Chord silence ratio within each beat of the generation.}
    \label{fig:metrics_silence}
    \end{center}
\end{figure}

To further investigate the \emph{wait and see} behavior noted at the end of \S\ref{sec:adaptation_measure},
we measure ratio of chord silence during a non-silence note across each beat.
Figure~\ref{fig:metrics_silence} shows that the finetuned models \RLChords and \RLChords-M,
are often silent during the first few beats.
We believe this behavior results from trading off the penalty for silence and the low reward for bad guesses.

\section{Knowledge Distillation Schedules}\label{sec:appendix_kd}

On-policy knowledge distillation~\cite{agarwal2023gkd,zhou2023distillspec} allows the KL to be evaluated with respect to any mixture of policy, teacher and data distribution.
We compare these options and report results in Table~\ref{tab:knowledge_distillation_metrics}.
Sampling from the policy alone results in good performance, and we use this choice in all of our experiments.

\section{Details of Regularization Penalties}
RL finetuning can lead to pathological behavior, such as repetitions and mode collapse~\cite{jaques2017sequence,jiang2020rl}.
We introduce several regularization penalties to discourage specific failure modes.

\textbf{Repetition} \space\space The model has a tendency to repeat itself, generating long-held chords. Inspired by repetition penalties used for training language models (as in \citet{saleh2020hierarchical,jaques2020human}), we provide a penalty of $-1$ for each token in a chord that is held for longer than 32 frames (8 beats), because chords longer than 8 beats rarely occur in the training set.

\textbf{Silence} \space\space To avoid behavior where no chords are generated, we provide a penalty of $-1$ for each silence token if more than $4\%$ of frames are silent accompaniment to a non-silent input. This is the rate at which silent accompaniment occurs in the training set.
We omit this penalty for the first 8 frames (first half-note) to allow early adaptation.

\textbf{Ending early} \space\space To discourage early end-of-sequence (EOS) tokens, we provide a penalty of $-1$ for each token after model outputs EOS but before the input does.
The penalties are given to the whole sequence at the last token of output before EOS.
An ablation experiment in \S\ref{sec:appendix_penalties} demonstrates the need for these penalties.

\section{Ablation on Regularizers}\label{sec:appendix_penalties}
To show the need for regularization penalties, we conduct an ablation test. As shown in the Table~\ref{tab:penalty_metrics}, removing the repetition penalty results in significantly more long chords. Removing the silence penalty would result in a long chunk of silence at the beginning of the generation, while removing the early-finish penalty would result in high ratio of samples stop early.

\begin{table}[t]
\caption{Comparing different knowledge distillation schedule, generating with policy data achieves best balance between harmonic, synchronization and rhythmic diversity metrics.}
\label{tab:knowledge_distillation_metrics}
\centering
\begin{tabular}{@{}lrrr@{}}
\toprule
Models                     & \multicolumn{1}{c}{\begin{tabular}[c]{@{}c@{}}Melody Note in Chord Ratio\\ (harmony) $\uparrow$\end{tabular}} & \multicolumn{1}{c}{\begin{tabular}[c]{@{}c@{}}$\Delta$ Chord-Note Onset Interval\\ $\times 10^{-3}$ (synchronization) $\downarrow$\end{tabular}} & \multicolumn{1}{c}{\begin{tabular}[c]{@{}c@{}}Chord Length Entropy\\ (rhythm diversity) $\uparrow$\end{tabular}} \\ \midrule
Online MLE                 & 36.99                                                                                              & 14.23                                                                                                            & 2.21                                                                                                  \\ \midrule
Policy (KD)                     & 46.54                                                                                              & 12.81                                                                                                            & 1.79                                                                                                  \\
Dataset                    & 35.25                                                                                              & 19.58                                                                                                            & 2.05                                                                                                  \\
Dataset + Policy           & 46.28                                                                                              & 13.65                                                                                                            & 1.67                                                                                                  \\
Dataset + Policy + Teacher & 46.64                                                                                              & 13.39                                                                                                            & 1.63                                                                                                  \\ \midrule
Offline MLE                & 63.73                                                                                              & 9.85                                                                                                             & 1.90                                                                                                  \\
Test set                   & 70.94                                                                                              & 0                                                                                                                & 2.19                                                                                                  \\ \bottomrule
\end{tabular}
\end{table}

\begin{table}[t!]
\caption{Comparing quantitative metrics of ablation experiments on regularization penalties, the proposed penalties effectively prevent RL training to exploit reward which result in sub-optimal solution. Details in \S\ref{sec:appendix_penalties}.}
\label{tab:penalty_metrics}
\begin{tabular}{@{}lrrrrrr@{}}
\toprule
\multicolumn{1}{c}{Models}                                                                        & \multicolumn{1}{c}{\begin{tabular}[c]{@{}c@{}}Note-in-Chord\\ Ratio (\%)\end{tabular}} & \multicolumn{1}{c}{\begin{tabular}[c]{@{}c@{}}Chord-Note\\ Onset Interval\end{tabular}} & \multicolumn{1}{c}{\begin{tabular}[c]{@{}c@{}}Chord Length \\ Entropy\end{tabular}} & \multicolumn{1}{c}{\begin{tabular}[c]{@{}c@{}}Chord Silence\\ Ratio (\%)\end{tabular}} & \multicolumn{1}{c}{\begin{tabular}[c]{@{}c@{}}Long Chords \\ Ratio (\%)\end{tabular}} & \multicolumn{1}{c}{\begin{tabular}[c]{@{}c@{}}Early Stop\\ Ratio (\%)\end{tabular}} \\ \midrule
Online MLE                                                                                        & 36.99                                                                                  & 14.23                                                                                   & 2.21                                                                                & 7.31                                                                                   & 1.67                                                                                  & 21.66                                                                          \\
Offline MLE                                                                                       & 63.73                                                                                  & 9.85                                                                                    & 1.90                                                                                & 2.91                                                                                   & 0.73                                                                                  & 2.62                                                                           \\
Test set                                                                                          & 70.94                                                                                  & 0                                                                                       & 2.19                                                                                & 2.83                                                                                   & 0.71                                                                                  & 0                                                                              \\
KD                                                                                                & 46.54                                                                                  & 12.8                                                                                    & 11.79                                                                               & 4.55                                                                                   & 1.51                                                                                  & 0                                                                              \\
ReaLchords                                                                                        & 48.17                                                                                  & 13.01                                                                                   & 1.25                                                                                & 4.90                                                                                   & 0.02                                                                                  & 0                                                                              \\ \midrule
\begin{tabular}[c]{@{}l@{}}KD\\ w/o all penalties\end{tabular}                                    & 45.96                                                                                  & 14.98                                                                                   & 1.66                                                                                & 32.67                                                                                  & 0                                                                                     & 0.38                                                                           \\ \midrule
\begin{tabular}[c]{@{}l@{}}ReaLchords\\ w/o repetition penalty\end{tabular}                       & 48.19                                                                                  & 13.17                                                                                   & 1.31                                                                                & 4.81                                                                                   & 0.78                                                                                  & 0                                                                              \\ \midrule
\begin{tabular}[c]{@{}l@{}}ReaLchords\\ w/o repetition penalty\\ w/o silence penalty\end{tabular} & 46.38                                                                                  & 11.12                                                                                   & 1.43                                                                                & 26.50                                                                                  & 0.32                                                                                  & 0                                                                              \\ \midrule
\begin{tabular}[c]{@{}l@{}}ReaLchords\\ w/o all penalties\end{tabular}                            & 46.92                                                                                  & 11.55                                                                                   & 1.46                                                                                & 27.85                                                                                  & 0.29                                                                                  & 0                                                                              \\ \bottomrule
\end{tabular}
\end{table}

\section{Training and Architecture Details of Online and Offline Models}

The online model $\Ponline_\theta$ is implemented as an 8-layer autoregressive decoder-only transformer~\cite{vaswani2017attention} with 6 heads and hidden dimension of 512.
It is trained on interleaved input and output tokens $y_1,x_1,\ldots,y_T,x_T$,
providing conditional distributions of the form $\pi_\theta(y_t\mid \pastxy)$ and $\pi_\theta(x_t\mid x_{<t},y_{\leqslant t})$.
To use this model \emph{online} as per \eqref{eqn:online},
we alternate drawing $y_t$s from $\pi_\theta(y_t\mid \pastxy)$ and $x_t$s from the given melody, foregoing the use of the conditionals $\pi_\theta(x_t\mid x_{<t},y_{\leqslant t})$. The online model is trained using Adafactor optimizer~\cite{shazeer2018adafactor} and learning rate of $10^{-3}$ with a batch size of 256. The online model is trained for $50,000$ steps with $1000$ steps of warmup. We apply a dropout with rate $0.1$ to the online model during training.

The offline model $\Poffline_\omega$ is implemented as an 8-layer encoder-decoder transformer in the style of T5~\cite{raffel2020exploring} with 6 heads and hidden dimension of 512.
This model is trained on pairs of sequences $x,y$ without interleaving, by first encoding the entire melody $x$ and then generating the entire chord progression $y\mid x$ conditioned on the encoding of $x$. The offline model is trained using Adafactor optimizer~\cite{shazeer2018adafactor} and learning rate of $10^{-3}$ with a batch size of 256. The offline model is trained for $50,000$ steps with $1000$ steps of warmup. We apply a dropout with rate $0.1$ to the offline model during training. We train our online and offline transformers using the T5X framework~\cite{roberts2023scaling}.

\section{Training and Architecture Details of Reward Models}\label{sec:appendix_reward_models}

The training specification is identical for reward models at all scale, with the only difference lies in their input length. Our contrastive reward model consists of two identical 6-layer, 6-head transformer encoders with 512 hidden dimension, one for encoding the entire melody $x$ and one for encoding the entire chord progression $y$. The contrastive reward model is trained using Adam optimizer~\cite{kingma2014adam} and learning rate of $10^{-4}$ with a batch size of 128. The contrastive reward model is trained for 35 epochs with a dropout rate of $0.1$.

The discriminative reward model consists of a 6-layer, 6-head transformer encoder with 512 hidden dimension that encodes the entire composition $x,y$.The discriminative reward model is trained using Adam optimizer~\cite{kingma2014adam} and learning rate of $10^{-4}$ with a batch size of 64. The discriminative reward model is trained for 5 epochs which we find after which point the model starts to overfit. We also applied a dropout rate of $0.1$ during training.

\section{Test Set performance of Reward Models}\label{sec:appendix_reward_model_performance}

We evaluate the contrastive and discriminative reward models on test set, and report performance in Table~\ref{tab:contrastive_model} for contrastive model, and Table~\ref{tab:discriminative_model} for discriminative model. For contrastive model, we evaluate it on retrieval metrics, namely recall (R@1, R@5, R@10) and mean average precision (mAP@10). For discriminative model, we use test set pair as positive label, and generate same number of negative samples by taking randomly pairing across test set. While performance of discriminative model remains similar across all scale, for contrastive model, the retrieval performance degrades at smaller scale. Given the focus of contrastive models on harmony evaluation, we posit that the observed decline in performance at smaller scales is attributed to the increased chance of harmonic coherence between unmatched samples in shorter segments.

\begin{figure}[ht]%
\centering
\includegraphics[width=0.5\linewidth]{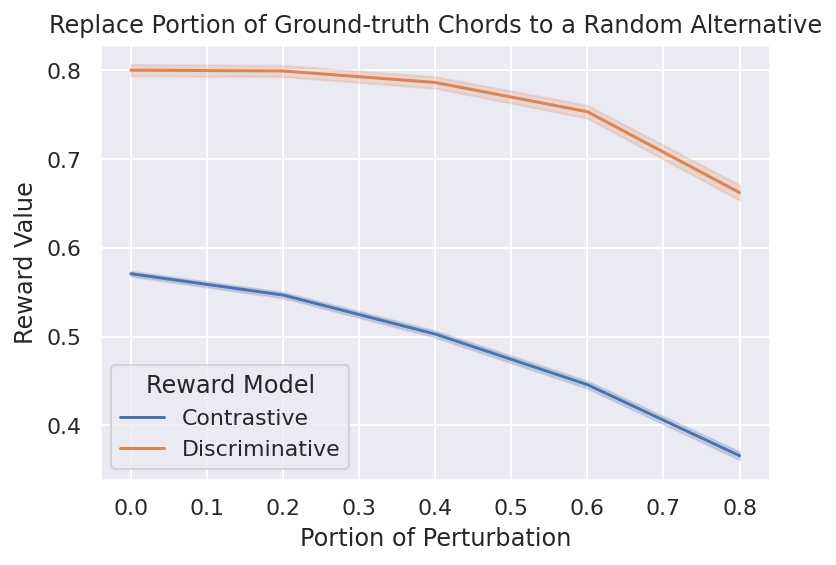}
\caption{We take ground-truth samples from the test set, and replace a portion of chords with random chord names but keep the chord boundary. We then run the perturbed sample through contrastive and discriminative reward models. The points in line show the average reward value while shaded area shows the confidence interval (95\%).The results show that the contrastive model is more attentive to harmonic perturbation or ``mistakes". Both reward models lower their scores as perturbation or ``mistakes" increases.}
\label{fig:reward_perturb}
\end{figure}

We examine the bias of different reward models by plotting reward values against harmonic perturbations, where varying portions of chords in the test set are replaced with random alternatives. As shown in Figure~\ref{fig:reward_perturb}, the contrastive model is more sensitive to harmonic perturbations.    

\section{Training Details of RL Finetuning}

In RL training, we use a learning rate of $10^{-4}$ for both policy model and value model. In the initialization of RL finetuning, the policy model and the value model are both initialized from online MLE model. The coefficient $\beta$ between reward maximization and KL loss in Equation~\ref{eqn:objective} is fixed as $0.5$ for all the experiments. We apply a coefficient of 50 to the reward produced by reward models. We apply a coefficient of 20 to the ending early penalty in all experiments used this penalty. In experiment using only knowledge distillation objective or using only one reward model, we apply a coefficient of 1 to both the repetition penalty and silence penalty. To train \RLChords, we apply a coefficient of 2 to both the repetition penalty and silence penalty. To train \RLChords-M, we apply a coefficient of 10 to both the repetition penalty and silence penalty.

We do not backpropagate gradients through the sampling process of the policy. All rewards, including reward models and penalties are summed and applied to the last token policy generated. For knowledge distillation objective, we only backpropagate gradients through the tokens policy generated, excluding the input tokens.

For systems using only reward models (C, D, C+D in Table~\ref{tab:main_result}), in addition to regularization penalties, we also include the knowledge distillation loss between policy and online MLE model. Similar to \citet{jaques2017sequence}, we found without such KL regularization, the training will be unstable and the models will generate invalid token sequences (e.g. a hold token before any onset token).

We also attempt to reproduce similar methods used in previous work RL-Duet~\cite{jiang2020rl} but received little success. We explored using offline MLE model as reward model, or using both offline MLE model and online MLE model as reward models, alongside with all regularization penalties we use. In all cases, the trained policy only generates repeated same chord, failing to generate adequate accompaniment.

\begin{table}[t]
\caption{The retrieval performance of contrastive reward model on test set.}
\label{tab:contrastive_model}
\centering
\begin{tabular}{lcccccccc}
\hline
\multirow{2}{*}{Contrastive Reward Models} & \multicolumn{4}{c}{Note to Chord}                & \multicolumn{4}{c}{Chord to Note} \\ \cline{2-9} 
                                           & R@1  & R@5  & R@10 & \multicolumn{1}{c|}{mAP@10} & R@1    & R@5    & R@10  & mAP@10  \\ \hline
Full context (256 frames)                  & 0.17 & 0.39 & 0.49 & \multicolumn{1}{c|}{0.26}   & 0.17   & 0.39   & 0.51  & 0.27    \\
1/2 context (128 frames)                   & 0.05 & 0.14 & 0.21 & \multicolumn{1}{c|}{0.09}   & 0.05   & 0.15   & 0.21  & 0.09    \\
1/4 context (64 frames)                    & 0.02 & 0.08 & 0.13 & \multicolumn{1}{c|}{0.05}   & 0.02   & 0.07   & 0.12  & 0.05    \\
1/8 context (32 frames)                    & 0.02 & 0.06 & 0.10 & \multicolumn{1}{c|}{0.04}   & 0.02   & 0.05   & 0.09  & 0.03    \\
1/16 context (16 frames)                   & 0.01 & 0.04 & 0.07 & \multicolumn{1}{c|}{0.03}   & 0.01   & 0.03   & 0.06  & 0.02    \\ \hline
\end{tabular}
\end{table}

\begin{table}[t]
\caption{The classification performance of discriminative reward model on test set.}
\label{tab:discriminative_model}
\centering
\begin{tabular}{@{}lccc@{}}
\toprule
Discriminative Reward Models & Precision & Recall & F1   \\ \midrule
Full context (256 frames)    & 0.69      & 0.91   & 0.79 \\
$1/2$ context (128 frames)         & 0.69      & 0.92   & 0.79 \\
$1/4$ context (64 frames)          & 0.71      & 0.84   & 0.77 \\
$1/8$ context (32 frames)          & 0.69      & 0.88   & 0.77 \\
$1/16$ context (16 frames)          & 0.68      & 0.79   & 0.73 \\ \bottomrule
\end{tabular}
\end{table}

\section{Ensemble Same Type of Reward Models}\label{sec:appendix_ensemble_reward}

We conduct experiments on RL finetuning with ensemble of same type of reward models, with results shown in Table~\ref{tab:ensemble_reward}. For both contrastive and discriminative model, ensemble with same kind helps improving better metric value.

\begin{table}[h]
\caption{Effect of RL finetuning with our reward models and knowledge distillation on harmonic, synchronization and rhythmic diversity metrics. In row 2-7, we report confidence interval (95\%) of metric values over 3 trainings, each with different random seeds. We additionally include results of RL finetuning with C+C and D+D (two contrastive or discriminative models trained with different initial seeds). From the metrics value, the ensemble of reward models helps achieving better metric performances.}
\label{tab:ensemble_reward}
\centering
\begin{tabular}{@{}lccc@{}}
\toprule
Models                      & \multicolumn{1}{c}{\begin{tabular}[c]{@{}c@{}}Harmony \\ Note in Chord $\uparrow$, \%\end{tabular}} & \multicolumn{1}{c}{\begin{tabular}[c]{@{}c@{}}Synchronization \\ $\Delta$ Chord-Note Onset Interval $\downarrow$, $\times 10^{-3}$ \end{tabular}} & \multicolumn{1}{c}{\begin{tabular}[c]{@{}c@{}}Rhythm Diversity \\ Chord Length Entropy $\uparrow$\end{tabular}} \\ \midrule
Online MLE                  & 36.99                                                                                              & 14.23                                                                                                            & 2.21                                                                                                  \\ \midrule
Knowledge Distillation (KD) & 46.38 $\pm$ 0.45                                                                                               & 14.39 $\pm$ 1.25                                                                                                            & {1.80} $\pm$ 0.04                                                                                     \\
Contrastive (C)             & 47.22 $\pm$  1.22                                                                                           & 16.61 $\pm$ 4.78                                                                                                           & 1.80  $\pm$  0.29                                                                                              \\
Discriminative (D)          & 44.29 $\pm$ 0.99                                                                                              & 12.34 $\pm$ 7.95                                                                                                            & 1.70  $\pm$ 0.06                                                                                               \\
C + D                       & 46.12 $\pm$ 0.95                                                                                            & 12.86 $\pm$ 6.17                                                                                                & 1.56 $\pm$ 0.05                                                                                                \\
\RLChords                 & 48.17 $\pm$ 0.27                                                                                              & 16.09 $\pm$ 3.63                                                                                                          & 1.35 $\pm$ 0.30                                                                                                \\
\RLChords-M                  & {54.29} $\pm$ 1.55                                                                                   & 17.17 $\pm$ 4.86                                                                                                          & 1.66 $\pm$ 0.20                                                                                                \\ \midrule
C + C                  & 47.30 $\pm$ 1.63     & 10.13 $\pm$ 0.44        & 1.55 $\pm$ 0.05 \\
D + D                  & 46.15 $\pm$ 0.92  & 10.56 $\pm$ 1.33  & 1.63  $\pm$ 0.18                                                                                             \\\midrule
Offline MLE                 & 63.73                                                                                              & \phantom{0}9.85                                                                                                             & 1.90                                                                                                  \\
Test set                    & 70.94                                                                                              & \phantom{0}0.00                                                                                                                & 2.19                                                                                                  \\ \bottomrule
\end{tabular}
\end{table}

\section{Sample preparation for human evaluations}\label{sec:appendix_human}
The listening test consisted of pairwise comparisons between different systems. The pairwise comparisons are prepared as following: We randomly sample 32 melodies from the test set, and have each system generate an accompaniment to the first eight bars of a song. For each melody, the accompaniments from each system is paired with all other systems, yielding $\binom{4}{2} = 6$ pairs. Going through all 32 melodies, we obtain $6 * 32 = 192$ total pairs. We recruited ten musicians, who each rated 15 to 20 pairs.

\section{Quantitative Metrics Details}

The entropy of chord length uses a base of $e$ result in unit of nats. The EMD of chord-to-note onset interval reported in Table~\ref{tab:main_result} is multiplied by $10^3$ for better comparison. To determine the distribution of chord-note onset intervals, we categorized these intervals into bins defined by the number of frames, using bin boundaries set at $[0, 1, 2, ..., 16, 17, \infty]$. Similarly, for the distribution of chord lengths, we organized the data into bins based on the number of frames, with bin boundaries established at $[0, 1, 2, ..., 32, 33, \infty]$. Following this, we generated histograms for both the chord-note onset intervals and chord lengths. Finally, we computed the Earth Mover's Distance (EMD) between these histograms.

For note-in-chord ratio of a song, we average across the binary value at each frame representing whether the note is in a chord. For calculating whether the note is in a chord, we exclude the frames where either input melody or output chord is silence. Then, to report the note-in-chord ratio of a system, we average across the note-in-chord ratio of each song. To report the note-in-chord ratio at certain beat reported in \S\ref{sec:adaptation_measure} and \S\ref{sec:appendix_silence}, for each song we only consider the frames at that beat, and when averaging across songs, we exclude the song where the whole beat is silence.

\section{Dataset and Data Representation Details}\label{sec:appendix_dataset}

\begin{figure}[h]
    \begin{center}
    \includegraphics[width=0.5\linewidth]{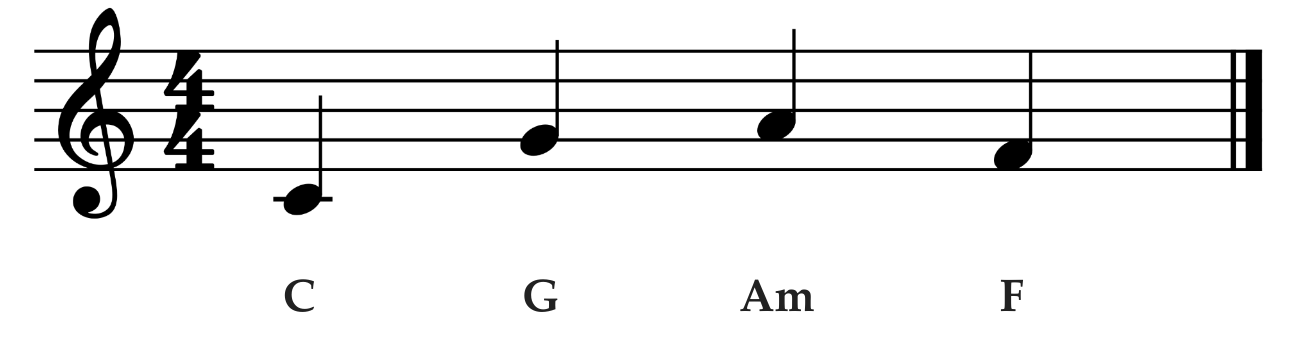}
    \caption{A visualization of example data samples.}
    \label{fig:data_example}
    \end{center}
\end{figure}

In \RLChords, we represent each chord by a unique name, assigning each distinct chord name a unique index. Melody tokens $x_t$ indicate both pitch and whether they mark the onset of a new note or the continuation of the previous token. Similarly, chord tokens $y_t$ denote the chord symbol and its onset or continuation status. The Hooktheory dataset we use comprises 5041 distinct chords. For the output $y$, each frame $y_t$ can be one of three possibilities: a chord selected from the 5041 available options, a chord hold (indicating the continuation of the previous chord) from the same set of options, or a silence, which signifies the generation of silence in the current frame. Similarly, there are 128 possible melody pitch values, ranging from 0 to 127 according to the MIDI pitch standard.
For the input $x$, each frame $x_t$ also falls into one of three categories: a note-on event chosen from 128 possible values, a note hold (indicating the continuation of the previous note) among the same 128 values, or a silence, indicating no input for the current frame.

As an example shown in Figure~\ref{fig:data_example}, the input melody $x$ for that example at each frame would be: [C4\_on, C4\_hold, C4\_hold, C4\_hold, G4\_on, G4\_hold, G4\_hold, G4\_hold, A4\_on, A4\_hold, A4\_hold, A4\_hold, F4\_on, F4\_hold, F4\_hold, F4\_hold], and the output chords $y$ for that example at each frame would be: [C\_on, C\_hold, C\_hold, C\_hold, G\_on, G\_hold, G\_hold, G\_hold, Am\_on, Am\_hold, Am\_hold, Am\_hold, F\_on, F\_hold, F\_hold, F\_hold].

The maximum length of the examples used for training is set to 256 frames, equivalent to 64 beats, with any samples exceeding this duration being randomly cropped during training.

\section{Histogram Statistics of Compared Systems}\label{sec:appendix_histograms}

\begin{figure}[h]
    \begin{center}
    \begin{subfigure}{\textwidth}
    \includegraphics[width=\linewidth]{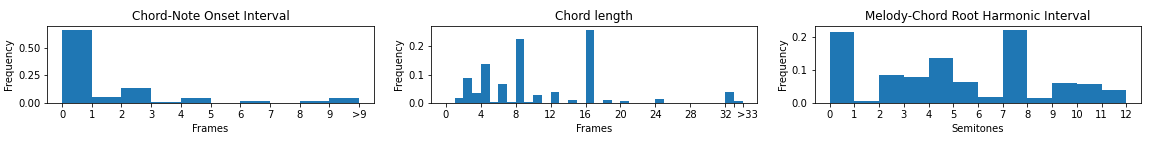}
    \caption{Test Set}
    \label{fig:histogram_ground_truth}
    \end{subfigure}
    \begin{subfigure}{\textwidth}
    \includegraphics[width=\linewidth]{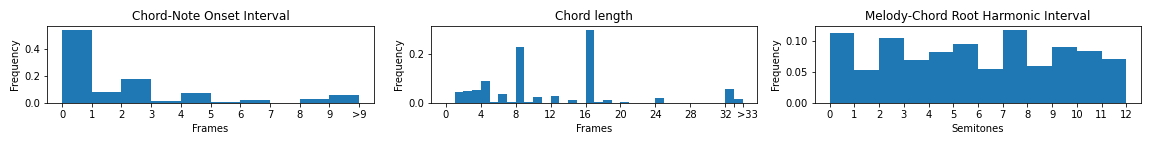}
    \caption{Online MLE}
    \label{fig:histogram_online}
    \end{subfigure}
    \begin{subfigure}{\textwidth}
    \includegraphics[width=\linewidth]{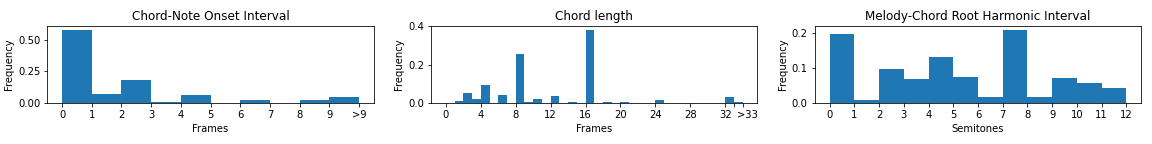}
    \caption{Offline MLE}
    \label{fig:histogram_offline}
    \end{subfigure}
    \begin{subfigure}{\textwidth}
    \includegraphics[width=\linewidth]{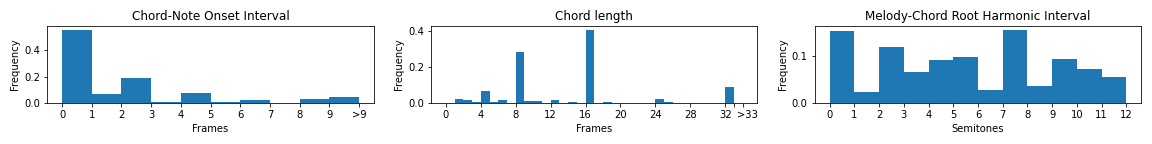}
    \caption{Knowledge Distillation (KD)}
    \label{fig:histogram_kd}
    \end{subfigure}
    \begin{subfigure}{\textwidth}
    \includegraphics[width=\linewidth]{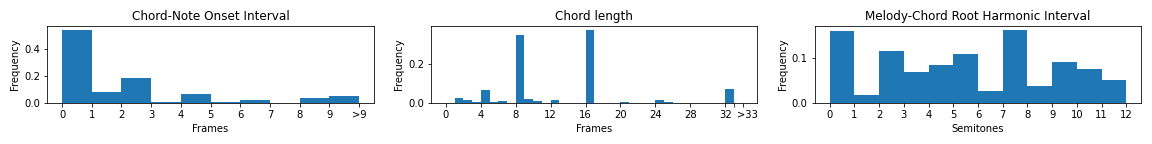}
    \caption{Contrastive Reward Model (C)}
    \label{fig:histogram_contrastive}
    \end{subfigure}
    \begin{subfigure}{\textwidth}
    \includegraphics[width=\linewidth]{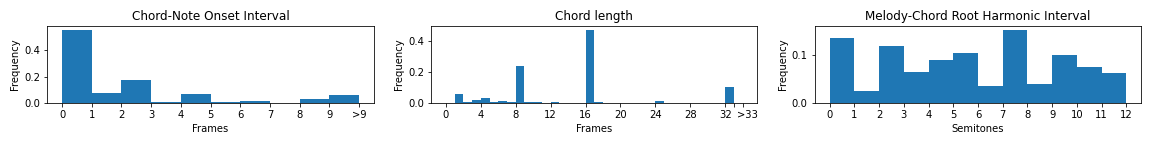}
    \caption{Discriminative Reward Model (D)}
    \label{fig:histogram_discriminative}
    \end{subfigure}
    \caption{The histogram of chord onset interval, chord length and harmonic interval for each system compared in \S\ref{sec:quantitative_results}.}
    \label{fig:histograms}
    \end{center}
\end{figure}

\begin{figure}[t!]
    \begin{center}
    \begin{subfigure}{\textwidth}
    \includegraphics[width=\linewidth]{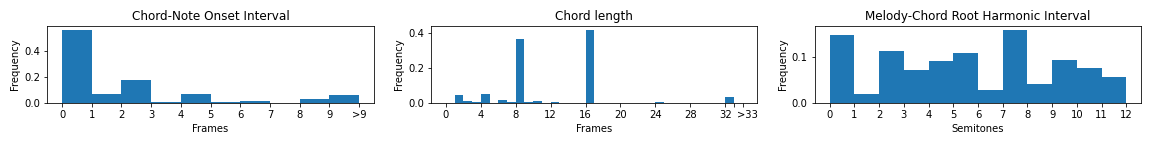}
    \caption{Contrastive + Discriminative Reward Model (C+D)}
    \label{fig:histogram_contrastive_discriminative}
    \end{subfigure}
    \begin{subfigure}{\textwidth}
    \includegraphics[width=\linewidth]{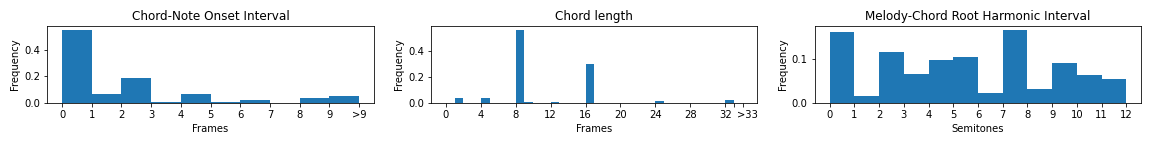}
    \caption{\RLChords}
    \label{fig:histogram_kd_c_d}
    \end{subfigure}
    \begin{subfigure}{\textwidth}
    \includegraphics[width=\linewidth]{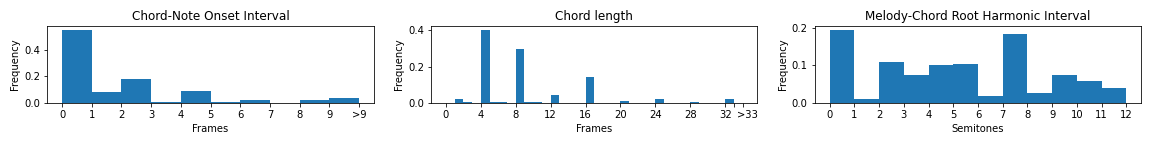}
    \caption{\RLChords-M}
    \label{fig:histogram_realchords}
    \end{subfigure}
    \caption{The histogram of chord onset interval, chord length and harmonic interval for each system compared in \S\ref{sec:quantitative_results}.}
    \label{fig:histograms_2}
        \end{center}
\end{figure}

In Figure~\ref{fig:histograms} and Figure~\ref{fig:histograms_2}, we present the histogram of chord-to-note onset interval, chord length and note-to-chord harmonic interval for one of each system compared in Table~\ref{tab:main_result}. Those histograms are used to calculate statistics report in Table~\ref{tab:main_result}. 

\end{document}